\documentclass[aps,prl,twocolumn,superscriptaddress,showpacs]{revtex4-1}
\usepackage{graphicx}
\usepackage{amssymb}
\usepackage{amsmath}
\pdfoutput=1

\begin{document}

\title{Electronic Pumping of Quasiequilibrium Bose-Einstein Condensed Magnons }

\author{Scott A. Bender}
\affiliation{Department of Physics and Astronomy, University of California, Los Angeles, California 90095, USA}
\author{Rembert A. Duine}
\affiliation{Institute for Theoretical Physics, Utrecht University, Leuvenlaan 4, 3584 CE Utrecht, The Netherlands}
\author{Yaroslav Tserkovnyak}
\affiliation{Department of Physics and Astronomy, University of California, Los Angeles, California 90095, USA}

\date{\today}

\begin{abstract}
We theoretically investigate spin transfer between a system of quasiequilibrated Bose-Einstein condensed magnons in an insulator in direct contact with a conductor. While charge transfer is prohibited across the interface, spin transport arises from the exchange coupling between insulator and conductor spins. In normal insulator phase, spin transport is governed solely by the presence of thermal and spin-diffusive gradients; the presence of Bose-Einstein condensation (BEC), meanwhile, gives rise to a temperature-independent condensate spin current.  Depending on the thermodynamic bias of the system, spin may flow  in either direction across the interface, engendering the possibility of a dynamical phase transition of magnons. We discuss experimental feasibility of observing a BEC steady state (fomented by a spin Seebeck effect), which is contrasted to the more familiar spin-transfer induced classical instabilities.
\end{abstract}


\maketitle

Bose-Einstein condensation (BEC) has been observed in a growing number of physical systems including trapped ultracold atoms and molecules \cite{andersonSCI95,*mewesPRL96,*zwierleinPRL03,*zwierleinSCI06}, semiconductor exciton polaritons \cite{dengSCI02,*kasprzakNAT06,*baliliSCI07}, and microcavity photons \cite{klaersNAT10}. In magnetic insulators, a quasiequilibrated BEC of magnons was created at room temperature by parametric pumping \cite{demokritovNAT06,*demidovPRL08}, which is especially intriguing as it represents the possibility of phase transitions in spintronic devices. In the case of short-lived bosonic excitations such as polaritons, photons, and magnons, the system needs to be optically pumped to exhibit spontaneous condensation \cite{snokeNAT06}.

In magnetic systems, Gilbert damping of magnons is known to increase upon the introduction of an adjacent conductor \cite{tserkovPRL02sp,*tserkovRMP05}: If the magnet is made to precess, conduction electrons may carry away spin upon colliding with the interface separating conductor and insulator, tilting the insulator's magnetization toward its axis of precession. Known as spin pumping, this magnetic relaxation process is reciprocal to spin-transfer torque \cite{slonczewskiJMMM96,bergerPRB96}, by which the angular momentum and energy can be pumped back into the magnetic region \cite{bauerPHYS11}. We consider here the consequences of these reciprocal interactions on an insulator with inhomogeneous spatial fluctuations in the magnetization, in particular a system of Bose-condensed magnons similar to that mentioned above. In this Letter, we construct rate equations for spin transfer between a magnetic insulator and adjacent normal metal, and solve for the time-dependent spin accumulation in the metal and the phase behavior of the insulator. The main text is supplemented with a discussion of the thermodynamics of spin transfer in our system and proposal of possible methods by which to detect the predicted dynamical phase transition.

\begin{figure}
\includegraphics[width=0.9\linewidth,clip=]{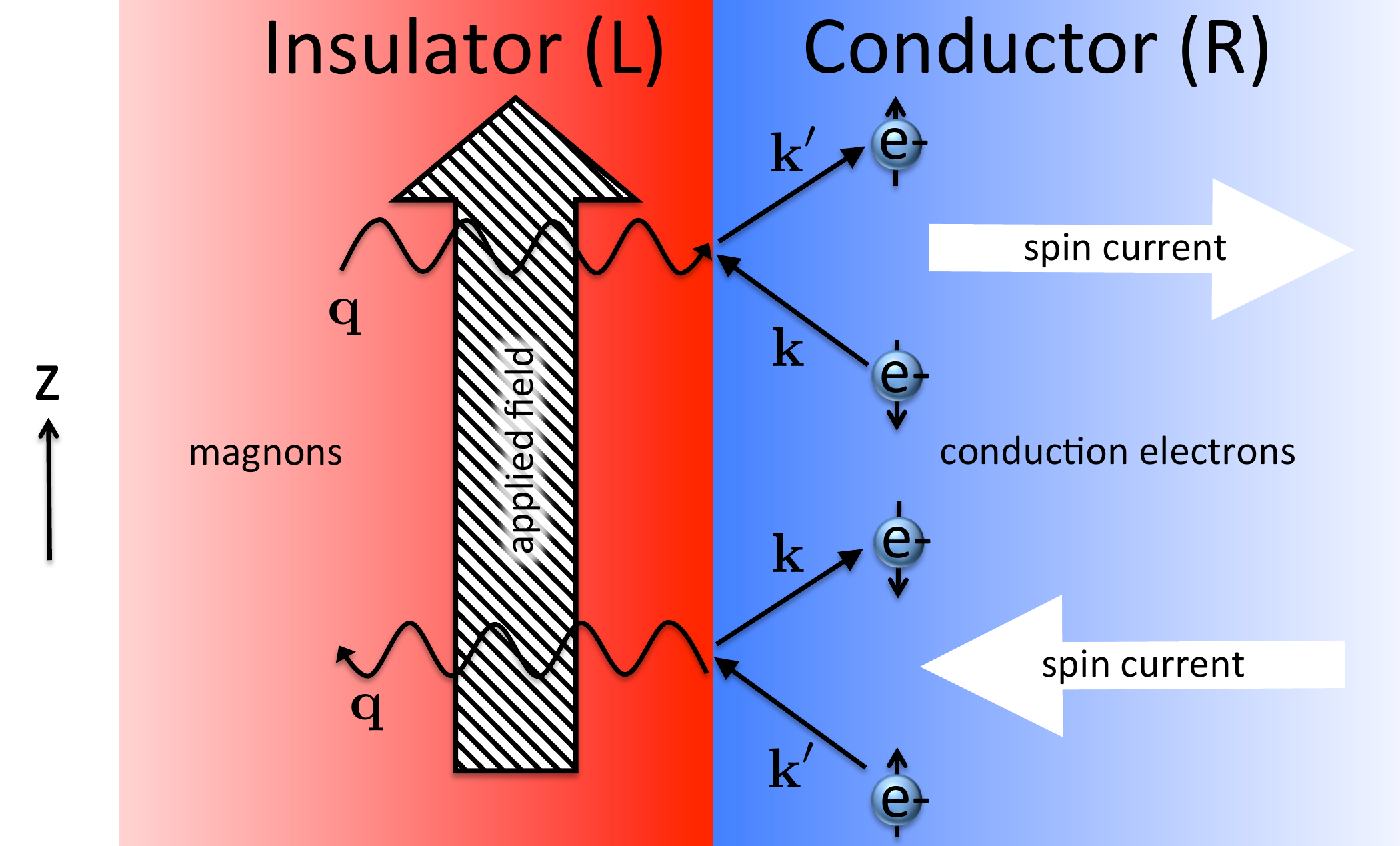}
\caption{The magnetic moments of insulator (left) atoms are coupled to the itinerant electrons of an adjacent conductor (right); an electron scatters inelastically off the interface, flipping its spin and creating or annihilating a magnon in the insulator.  While coupling across the interface requires some degree of overlap between electrons in the conductor and localized electron orbitals in the insulator, a net electron tunneling between the two subsystems is prohibited, so that only spin density is transferred.  The magnetic field in the insulator, and hence static magnetization, point in the positive $z$  direction; for a negative gyromagnetic ratio the static spin density is therefore oriented in the $-z$  direction, so that magnons carry spin $+\hbar$.}
\label{schematic}
\end{figure}

Let us consider the insulating ferromagnet subjected to a magnetic field $B$ in the positive
$z$ direction and attached to a metallic conductor, as sketched in Fig.~\ref{schematic}. Electrons in the ferromagnetic insulator are localized (typically in deep $d$ or $f$ orbitals) near atomics sites, precluding charge transport. The corresponding magnetic moments constitute individual degrees of freedom, which give rise to collective spin-wave excitations. Meanwhile, ($s$-character) electrons in the metal are considered completely delocalized and noninteracting. We shall henceforth
denote the ferromagnetic subsystem as {}``left'' or $L$, and the
metallic conductor subsystem as {}``right'' or $R$. As a starting point,
we treat them as uncoupled so that the electronic state of the entire system is $\left|m\right\rangle =\left|m_{L}\right\rangle \otimes\left|m_{R}\right\rangle $.
$\left|m_{L}\right\rangle $ is an eigenket of the linearized (i.e.,
noninteracting magnon) left Hamiltonian $\hat{H}_L$; in other words, it is
an element of the Fock space of Holstein-Primakoff (HP)
magnons, each indexed by the mode number $\mathbf{q}$. The magnon spectrum $\epsilon_{\mathbf{q}}$ is gapped [${\rm min}(\epsilon_{\mathbf{q}})=\epsilon_{\rm gs}>0$]
by the presence of the magnetic field or anisotropy. Meanwhile, $\left|m_{R}\right\rangle $ is an element of electron Fock space and represents an antisymmetrized product of single-particle states corresponding to quasiparticle Hamiltonian $\hat{H}_R$, each indexed by orbital quantum number $\mathbf{k}$  and spin $\sigma$.

Itinerant electrons in the conductor are coupled across the insulator-conductor
interface to the magnetic moments of the insulator by
a generic exchange interaction. We suppose that this interaction
$\hat{V}_{\mathrm{int}}$ can be phenomenologically written in terms
of creation (annihilation) operators $\hat{c}_{\mathbf{q}}^{\dagger}$
($\hat{c}_{\mathbf{q}}$) for free HP magnons and creation (annihilation) operators
$a_{\mathbf{k}\sigma}^{\dagger}$ ($a_{\mathbf{k}\sigma}$) for conduction electrons:
\begin{equation}
\hat{V}_{\mathrm{int}}=\sum_{\mathbf{q}\mathbf{k}\mathbf{k}'}V_{\mathbf{q}\mathbf{k}\mathbf{k}'}\hat{c}_{\mathbf{q}}\hat{a}_{\mathbf{k}'\uparrow}^{\dagger}\hat{a}_{\mathbf{k}\downarrow}+{\rm H.c.}\,,
\label{interaction}
\end{equation}
where $\sigma=\uparrow$ or $\downarrow $ denote electron spin in the $+z$ or $-z$ directions, respectively. Information about scattering off of the static component of the insulator magnetization is entirely contained in the conduction electron wavefunction $\psi_{\mathbf{k} \sigma} \left( \mathbf{x} \right)$, which we consider to have a finite albeit exponentially vanishing extension into the insulator; more specifically,  $\psi_{\mathbf{k} \sigma} \left( \mathbf{x} \right)$ are eigenstates of the total mean-field Hamiltonian, including the interaction just on the inside of the insulator between the evanescent conduction electron tails and the static $z$ component of the insulator magnetization.  We approximate the static component of the magnetization as spatially uniform in what follows. The effect on conduction electron scattering due to the \textit{rotating} magnetization component in the $xy$ plane, i.e., Eq.~($\ref{interaction}$), which we consider small in comparison to the static component, is responsible for spin pumping \cite{tserkovPRL02sp} and spin-transfer torque \cite{slonczewskiJMMM96,bergerPRB96} and treated perturbatively below.

The first term on the right-hand side of Eq.~(\ref{interaction}) describes a magnon (carrying spin up $\hbar$) annihilating in the insulator to create a spin-down hole/spin-up
electron pair in the conductor, while its Hermitian conjugate (H.c.) corresponds
to a reverse electron spin-flip scattering off the insulator-conductor
interface to create a magnon. The scattering amplitude $V_{\mathbf{q}\mathbf{k}\mathbf{k}'}$
is assumed to be a full matrix element describing this process. Notice that while energy is exchanged in this interaction,
momentum is not generally conserved. Moreover, this is not the only
means by which conduction electrons can exchange energy with the magnetic
insulator: One could, for example, write down an inelastic scattering term of the form $\sim \hat{c}_{\mathbf{q}'}^\dagger\hat{c}_{\mathbf{q}}\hat{a}_{\mathbf{k}'\sigma}^{\dagger}\hat{a}_{\mathbf{k}\sigma}$ that conserves magnon
number (and therefore preserves the spin of the scattering conduction
electron), which physically corresponds to a deviation of the spin-conserving part of the Hamiltonian from its mean-field form. Since such a process does not contribute to the transfer
of the $z$ component of spin across the interface, however, it becomes irrelevant when temperatures are maintained by thermal reservoirs.  It should also be noted that the presence of shape anisotropy generally gives rise to elliptical magnons. The elliptical magnon operators $\hat{b}_{\mathbf{q}}$ and $\hat{b}^{\dagger}_{\mathbf{q}}$ are linear combinations of circular magnon operators $\hat{c}_{\mathbf{q}}$ and $\hat{c}^{\dagger}_{\mathbf{q}}$, so that $\hat{c}_{\mathbf{q}}$ and $\hat{c}^{\dagger}_{\mathbf{q}}$ no longer diagonalize $\hat{H}_L$. While our detailed analysis in the following assumes circular magnons, a finite magnon eccentricity is not expected to significantly alter our findings qualitatively.

The total Hamiltonian can be expanded as $\hat{H}_{\mathrm{tot}}=\hat{H}_{L}+\hat{H}_{R}+\hat{V}_{\mathrm{int}}+\hat{H}_{T}+\hat{H}_{\mathrm{env}}$,
where $\hat{H}_T$
is a thermalizing Hamiltonian that contains magnon-magnon interactions
and conduction electron-electron interactions, while $\hat{H}_{\mathrm{env}}$
describes interactions between magnons and conduction electrons with their environments:
magnon-phonon coupling, electron-phonon coupling, etc.  Here we consider dephasing effects significant enough that coherence between the
left and right subsystems is destroyed and the density matrix 
for the entire system is always in the form $\hat{\rho}_{\mathrm{tot}}=\hat{\rho}_L\otimes\hat{\rho}_R$.  We further assert, subject to sufficiently fast thermalization in respective subsystems, that
\begin{align}
\mathrm{Tr}[\hat{\rho}_R\hat{a}_{\sigma \mathbf{k}}^{\dagger}\hat{a}_{\sigma'\mathbf{k}'}]&=n_F\left(\beta_{\mathrm{R}}(\epsilon_{\mathbf{k}}-\mu_{\sigma}) \right) \delta_{\mathbf{k}\mathbf{k}'}\delta_{\sigma\sigma'}\,,\nonumber\\
\mathrm{Tr}[\hat{\rho}_L\hat{c}_{\mathbf{q}}^{\dagger}\hat{c}_{\mathbf{q}'}]&=n_B\left( \beta_L(\epsilon_\mathbf{q}-\mu_{\mathrm{L}}) \right) \delta_{\mathbf{qq}'}\,,
\label{Lrestriction}
\end{align}
where $n_F(x)=(e^x+1)^{-1} $ and $n_B(x)=(e^x-1)^{-1} $ are the (quasiequilibrium) Fermi-Dirac and Bose-Einstein distributions, respectively, and $\epsilon_\mathbf{k}$ ($\epsilon_\mathbf{q}$) is the electron (magnon) spectrum.
Because each subsystem maintains internal equilibrium, magnons
obey Bose-Einstein statistics while conduction electrons are
described by a Fermi-Dirac distribution. Information about the allotment
of spin and energy between them is now contained in the inverse temperatures
$\beta_{L}$ and $\beta_{R}$, the chemical potential $\mu_{\sigma}$ for
conduction electrons with spin $\sigma$, and the effective magnon chemical
potential $\mu_{L}$ (which does not have to vanish in a pumped system). Note that $\mu_{L}\leq\epsilon_{\rm gs}$, where
$\epsilon_{\rm gs}$ is the ground-state magnon energy; the magnons
become Bose-Einstein condensed when $\mu_{L}=\epsilon_{\rm gs}$.

It is straightforward to calculate the spin current (per interfacial area $A$)  $j$
flowing into the insulator from the conductor in terms of temperatures
and chemical potentials to lowest order in $\hat{V}_{\mathrm{int}}$ using Fermi's golden rule:
\begin{equation}
j=\frac{1}{A}\frac{d\left\langle S_{L}^{z}\right\rangle }{dt}=j_{\rm gs}+j_{\rm ex},\label{eq: spin rate}\end{equation}
where the ground-state, $j_{\rm gs}$, and excited, $j_{\rm ex}$, magnon contributions are functions of the magnon chemical potential $\mu_L$, electron spin accumulation $\Delta \mu=\mu_\uparrow-\mu_\downarrow$, and their temperatures $T_L$ and $T_R$.
In the thermodynamic limit, the spin-current density $j_{\rm gs}$, describing the rate of flow of ground-state magnons into and out of the insulator, is proportional to the number of ground-state magnons $N_{\rm gs}$ per insulator volume $V_L$, $n_{\rm gs}=N_{\rm gs}(\mu_L,T_L)/V_L$: 
\begin{equation}
j_{\rm gs}=2\pi\left|V_{\rm gs}\right|^{2}\left(\Delta\mu-\epsilon_{\rm gs}\right)g_{R}^2n_{\rm gs}\,.
\label{eq: gs current}
\end{equation}
Here, $g_R$  is the Fermi-level density of states of conduction electrons and 
\begin{align}
\left|V_{\rm gs}\right|^{2}\equiv&\frac{V_L}{A}\left(\frac{V_{R}}{g_R}\right)^2\int\frac{d^{3}\mathbf{k}}{\left(2\pi\right)^3}\frac{d^{3}\mathbf{k}'}{\left(2\pi\right)^3}\left|V_{0\mathbf{k}'\mathbf{k}}\right|^{2}\nonumber\\
&\times\delta\left(\epsilon_{\mathbf{k}}-\epsilon_F\right)\delta\left(\epsilon_{\mathbf{k}'}-\epsilon_F\right)\,,
 \label{eq: v2BEC}
\end{align}
where $\epsilon_F$ is the Fermi energy (assumed to be much larger than $\epsilon_{\rm gs}$ and temperature) and $V_R$ volume of the conductor. Note that the current density $j_{\rm gs}$ is only present in the thermodynamic limit in BEC phase, $\mu_L=\epsilon_{\rm gs}$. On the other hand, the spin-current density $j_{\rm ex}$
(carrying spin transfer via the excited magnon states) is present in both normal and BEC phases and, after some manipulations, can be written as
\begin{align}
j_{\rm ex}=&2\pi\int_{\epsilon_{\rm gs}}^{\infty}d\epsilon\left|V_{\mathrm{ex}}(\epsilon)\right|^{2}\left(\Delta\mu-\epsilon\right)g_R^{2}g_L(\epsilon)\nonumber\\
&\times \left[n_{\mathrm{B}}\left(\beta_{L}(\epsilon-\mu_{L})\right) -n_{\mathrm{B}}\left(\beta_{R}(\epsilon-\Delta\mu)\right)\right]\,,
\label{eq: ex current}
\end{align}
in terms of the energy-dependent density of magnon states $g_L(\epsilon)$. The (relatively weakly) energy-dependent quantity
\begin{align}
\left|V_{\mathrm{ex}}(\epsilon)\right|^{2}\equiv&\frac{V_L}{Ag_L(\epsilon)}\left(\frac{V_R}{g_R}\right)^2\int\frac{d^{3}\mathbf{k}}{\left(2\pi\right)^3}\frac{d^{3}\mathbf{k}'}{\left(2\pi\right)^3}\frac{d^{3}\mathbf{q}}{\left(2\pi\right)^3}\left|V_{\mathbf{q}\mathbf{k}'\mathbf{k}}\right|^{2}
\nonumber\\
&\times\delta\left(\epsilon_{\mathbf{k}}-\epsilon_F\right)\delta\left(\epsilon_{\mathbf{k}'}-\epsilon_F\right)\delta\left(\epsilon_{\mathbf{q}}-\epsilon\right)
\end{align}
contains information about inelastic transition rates involving excited magnons.

The dynamics of spin flow across the interface are therefore determined
by the sum of the condensate current density $j_{\rm gs}$, which is determined by spin accumulation in the conductor and the ground-state magnon energy $\epsilon_{\rm gs}$ (and thus the applied magnetic field), and the thermal current density $j_{\rm ex}$, which depends on both temperature and spin-potential biases. Note that sufficiently large spin splitting $\Delta \mu$ in the conductor could, in principle, drive spin density into the insulator until the required density of magnons is attained and the system undergoes Bose-Einstein condensation. In a recent experiment by Sandweg \textit{et~al.} \cite{sandwegPRL11}, spin pumping into a metal by magnetic insulator is driven by the presence of parametrically excited magnons; in addition, a spin current between the metal and insulator arises from a thermal gradient as discussed above. The authors of Ref.~\cite{sandwegPRL11} made use of the inverse spin Hall effect, wherein spin diffusion along a metal strip produces detectable Hall signal. Reciprocally, an electric current could be used to generate spin accumulation on the surface of a metal via the spin Hall effect; this surface spin accumulation may then drive magnons into the insulator \cite{kajiwaraNAT10}. 

\begin{figure}
\includegraphics[width=0.8\linewidth,clip=]{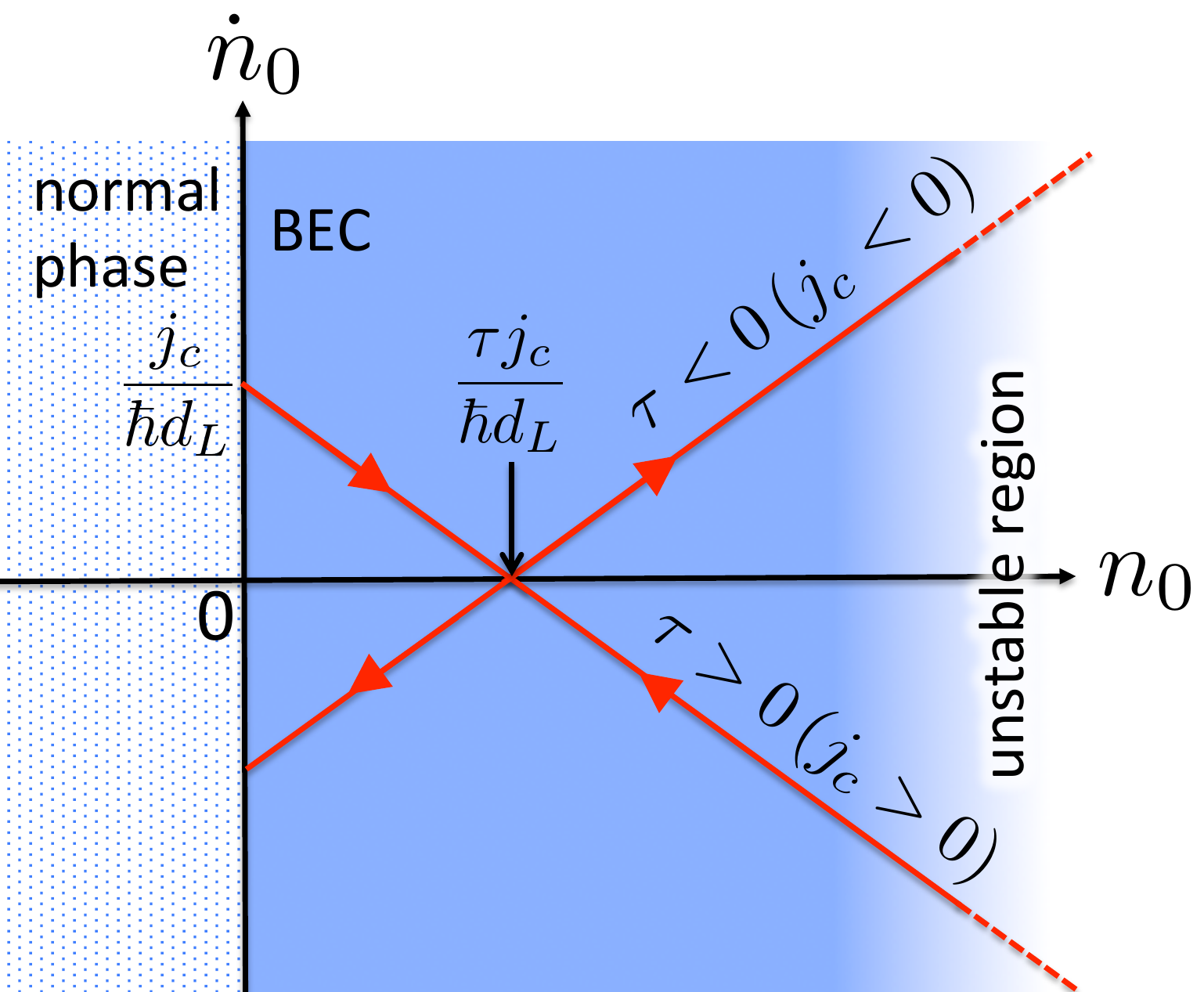}
\caption{Behavior of $n_{\mathrm{gs}}$ as predicted by the rate equation, $\dot{n}_{\mathrm{gs}}=j_{\mathrm{tot}}/\hbar d_L=j_c/\hbar d_L- n_{\mathrm{gs}}/\tau$. If $j_c$ had the sign opposite to that shown in the figure, the crossing point $\tau j_c/\hbar d_L$ would fall in the normal phase ($n_{\rm gs}=0$), thus precluding a BEC formation.}
\label{P}
\end{figure}

\begin{figure}
\includegraphics[width=0.9\linewidth,clip=]{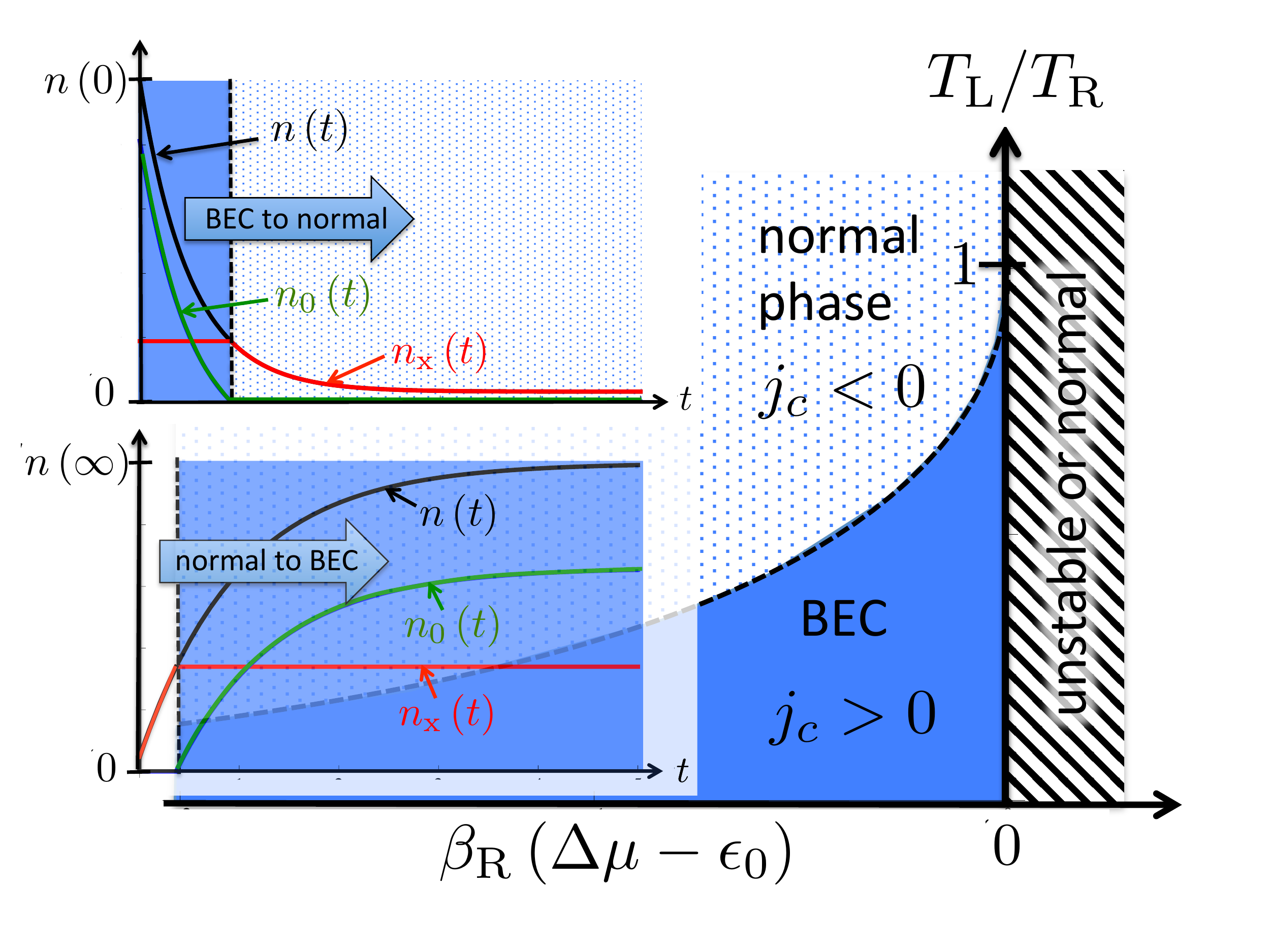}
\caption{When $\Delta \mu<\epsilon_{\rm gs}$, the steady-state phase is insensitive to the initial condition for $n_{\rm gs}$, but depends on the temperature bias $T_L-T_R$ and the difference $\Delta \mu-\epsilon_{\rm gs}$. As the splitting $\Delta \mu$ increases, the critical temperature for $T_L$ increases until it equals $T_R$. Examples of time dependence in the normal and BEC phase regions are shown in the upper and lower left panels, respectively. When $\Delta \mu>\epsilon_{\rm gs}$, depending on the initial condition, the driven magnon system is either unstable or relaxes towards the normal phase.}
\label{PB}
\end{figure}

We henceforth focus on the regime where the temperatures of both the left and right subsystems are fixed so that any energy gain or loss, independent of spin gain or loss, is completely absorbed or resupplied by thermal reservoirs. At fixed $T_L$ the density of excited magnons $n_{\rm ex}$ becomes a monotonic function of $\mu_L\leq\epsilon_{\rm gs}$  alone.   Let us further suppose that spin accumulation $\Delta \mu  $ in the right reservoir is independent of spin diffusion from the insulator and fixed. If the total density of magnons exceeds the critical BEC density $n_c$ (corresponding to $\mu_L=\epsilon_{\rm gs}$), $n_{\rm ex}$ reaches and remains pinned at this value, $n_c$, and only $n_{\rm gs}$ is free to vary. In BEC phase, then, the time dependence of $n_{\rm gs}$ is given by
\begin{equation}
n_{\rm gs}(t)=\frac{\tau j_c}{\hbar d_L}+\left[n_{\rm gs}(0)-\frac{\tau j_c}{\hbar d_L}\right]e^{-t/\tau}\,,
\label{ngs}
\end{equation}
where the excited magnon flux $j_c=j_{\rm ex}(\mu_L\to\epsilon_{\rm gs})$ is time independent, as long as $\mu_L$ is anchored by the condensate at $\epsilon_{\rm gs}$, $\hbar/\tau\equiv2\pi\left|V_{\rm gs}\right|^2\left(\epsilon_{\rm gs}-\Delta\mu\right)g_{R}^{2}/d_L$, and $d_L=V_L/A$ is the magnetic layer thickness. The behavior of the Bose-Einstein condensed system thus falls into one of four regimes, as depicted in Fig.~\ref{P}.  In the first, $\Delta \mu> \epsilon_{\rm gs}$ (so that $\tau^{-1}<0$) and $n_{\rm gs}(0)>\tau j_c/\hbar d_L$, $n_{\rm gs}$ grows exponentially until saturating at a  value $\sim M_s/\mu_B$ (where $M_s$ is the magnetization of the ferromagnet and $\mu_B$ is the Bohr magneton). In this case, magnon-magnon interactions become important ultimately and the system must be treated more carefully here. This is a realization of the ``swaser" (i.e., a spin-wave analog of a laser) put forward in Ref.~\cite{bergerPRB96} and observed in the context most similar to ours (in a magnetic insulator YIG) in Ref.~\cite{kajiwaraNAT10}. In the second regime, $\Delta \mu> \epsilon_{\rm gs}$ but $n_{\rm gs}(0)<\tau j_c/\hbar d_L$ (requiring $j_c<0$), $n_{\rm gs}$ decreases towards zero, and the system enters normal phase. The last two regimes (corresponding to $j_c>0$  and $j_c<0$), which are of more interest to us, occur when spin splitting in the conductor is sufficiently small that $\Delta \mu<\epsilon_{\rm gs}$ and thus $\tau^{-1}>0$, as depicted in Fig.~\ref{PB}. Here, the steady-state phase no longer depends on the initial condition: When $j_c>0$, the magnons will Bose-Einstein condense (lower half of the main panel in Fig.~\ref{PB}), and if $j_c<0$, normal phase with $n_{\rm gs}=0$ must eventually be reached (upper half of the main panel in Fig.~\ref{PB}).

In the normal phase ($n_{\rm ex}<n_c$), $\mu_L$  acquires time dependence, and the rate of change of the total number of magnons is $\dot{n}_{\mathrm{tot}}=\dot{n}_{\mathrm{ex}}=j_{\rm ex}(t)/\hbar d_L$. To illustrate these dynamics in a specific example,  we consider a simple model where the density of magnon states per unit insulator volume $V_{L}$ has the form $g_L(\epsilon)=\mathcal{G}_L(\epsilon/\epsilon_{\rm gs}-1)^w$ (with $w>0$ and $\mathcal{G}_L$ a positive real number).
In terms of the polylogarithm function
\begin{equation}
{\rm Li}_{w+1}\left(z\right)\equiv\frac{1}{\Gamma_{w+1}}\int_{0}^{\infty}dx\frac{x^w}{e^{x-\mathrm{ln}z}-1}\,,
\end{equation}
the density of excited magnons becomes
\begin{equation}
n_{\rm ex}=\eta^{\left(w\right)} (\beta_L,\mu_L) \equiv\mathcal{G}_L\frac{\Gamma_{w+1}{\rm Li}_{w+1}(z_{L})}{\beta_{L}^{w+1}\epsilon_{\rm gs}^w}\,,
\end{equation}
where $z_{L}(\beta_L,\mu_L)\equiv e^{\beta_{L}\left(\mu_{L}-\epsilon_{\rm gs}\right)}$
is the effective magnon fugacity (with $z_L=1$ corresponding to a BEC). Assuming for simplicity that $V_{\rm ex}(\epsilon)$ is energy independent and equal to $V_{\rm gs}$, one obtains from Eq.~\eqref{eq: ex current} an excited spin current
\begin{equation}
j_{\mathrm{ex}}=\frac{\hbar d_L}{\tau}  \left(\frac{\eta^{\left(w+1\right)}_R-\eta^{\left(w+1\right)}_L}{1-\Delta \mu/\epsilon_{\rm{gs}}}+\eta_R^{\left(w\right)}-\eta_L^{\left(w\right)}\right)\,,
\label{eq: excitedj}
\end{equation}
where $\eta^{\left(w\right)}_{L} \equiv  \eta^{\left(w\right)}(\beta_L,\mu_L)$ and $\eta^{\left(w\right)}_{R} \equiv  \eta^{\left(w\right)}(\beta_R,\Delta\mu)$. In general, to find the spin accumulation in the normal phase as a function of time, one must solve the rate equation for the magnon fugacity $z_L$. At low temperatures, $(\beta_L^{-1}, \beta_R^{-1})\ll|\epsilon_{\rm gs}-\Delta\mu|$, the first term in Eq.~\eqref{eq: excitedj} can be neglected, allowing for a simple solution to the excited magnon density:
\begin{equation}
n_{\mathrm{ex}}\left(t \right)=\eta_R^{\left(w \right)}+\left[ n_{\rm ex}(0)-\eta_R^{\left(w \right)} \right]e^{- t/\tau}\,,
\label{eq: ex(t)}
\end{equation}
provided $n_{\mathrm{ex}}<n_c$. If $\Delta \mu<\epsilon_{\rm gs}$, $\tau^{-1}>0$, and $n_{\rm ex}$ decays towards $\eta_R^{\left(w\right)}$, irrespective of its initial condition. If $\eta_R^{\left(w\right)}<n_{\mathrm{c}}$, the insulator always remains in normal phase; when $\eta_R^{\left(w \right)}>n_{\mathrm{c}}$, on the other hand, the magnons eventually Bose-Einstein condense, and the system is henceforth described by Eq.~(\ref{ngs}). Notice that the conditions $\eta_R^{\left(w\right)} \gtrless n_{\mathrm{c}}$ are (in the spirit of the aforementioned low-temperature approximation) equivalent to $j_c\gtrless 0$, which are consistent with the conditions considered above for the system to settle in the BEC or normal phase, respectively, as $t\rightarrow \infty$.  The time dependence in the opposite high-temperature regime, $\beta_L^{-1}, \beta_R^{-1}\gg|\epsilon_{\rm gs}-\Delta\mu|$, is more complicated than but in principle similar in behavior to the low-temperature solution given by Eq.~(\ref{eq: ex(t)}).

If the insulator temperature $T_L$ is left floating, the energy flow between the two subsystems would give rise to the dynamics of $T_L$ (supposing for simplicity $T_R$ is still fixed). In the most extreme case, the insulator is allowed to exchange energy only with the conductor (and only by the electron-magnon scattering discussed above, neglecting phonon heat transfer), so changes in $T_L$ are dictated by the rate at which energy is transferred across the barrier along with spin.
The coupled rate equations for energy and spin transfer can then
be solved to give time-dependent solutions to the temperature
$T_{L}$ and the ground and excited magnon densities, $n_{\rm ex}$ and $n_{\rm gs}$. While this program is beyond our scope here, we may expect a significantly more complex phase diagram, with hysteretic features sensitive to the initial conditions and reentrant phase behavior.

All of the relevant quantities may be readily inferred from existing measurements.  In particular, the squared matrix element $\left|V_{\rm{gs}} \right|^{2}$ is directly related to the real spin-mixing conductance (per unit area) $g^{\uparrow \downarrow}$ by equating the ground-state current density $j_{\rm gs}$ for $\Delta \mu=0$ with the expression for current pumped by a precessing magnetic monodomain given in Ref.~\cite{tserkovPRL02sp}: One obtains $\left|V_{\mathrm{gs}} \right|^{2}=g^{\uparrow \downarrow}/4\pi^2 s g_R^2$, where $s$ is the ferromagnetic spin density in units of $\hbar$. From this relation, the ``magnon dwell time" $\tau_d\equiv  \tau|_{\Delta \mu=0}=2\pi s d_L/g^{\uparrow \downarrow}\omega_r$ and the effective Gilbert damping constant $\alpha'\equiv1/2\omega_r\tau_d=g^{\uparrow\downarrow}/4\pi sd_L$ (corresponding to the interfacial, i.e., spin-pumping \cite{tserkovPRL02sp}, magnon decay) are expressed in terms of the spin-mixing conductance. ($\omega_r\equiv\epsilon_{\rm gs}/\hbar$ here is the ferromagnetic-resonance frequency.) We use the term ``Gilbert damping" here to refer to dynamical magnetization damping generally, including damping of inhomogeneous fluctuations, in lieu of the alternative ``Landau-Lifshitz" damping; while the two are mathematically equivalent, historically the former has become generally favored over the latter, and so we follow this convention. In YIG films ($4\pi M_s\approx2~\rm{kG}$, $g^{\uparrow \downarrow}\sim 10^{14}~\rm{cm}^{-2}$ \cite{kajiwaraNAT10,heinrichPRL11}), the spin-pumping Gilbert damping $\alpha'$ dominates over the intrinsic Gilbert damping ($\alpha\sim 10^{-4}$) below thicknesses $d_L\sim100$~nm. Theoretically predicted \cite{jia} and recently measured \cite{burrows} mixing conductance that is a factor of five larger ($g^{\uparrow \downarrow}\approx5\times10^{14}~\rm{cm}^{-2}$) proportionately increases the maximum film thickness. Having fixed $\alpha'$ for a given $d_L$, the applied magnetic field can be chosen to be sufficiently small that the timescale $\tau_{\rm{th}}$ for magnon thermalization is significantly less than the characteristic dwell time $\tau_d=1/2\alpha'\omega_r$. For example, taking $\tau_{\rm{th}}\sim100$~ns for room-temperature YIG \cite{demokritovNAT06,*demidovPRL08}, the dwell time $\tau\sim1$ $\mu$s for damping $\alpha'\sim10^{-4}$ corresponds to a frequency of $\sim100$~MHz or (effective) field of $\sim10$~G. At this field, the condition for the formation of BEC ($j_c>0$) requires a temperature bias $\Delta T=T_R-T_L\sim\epsilon_{\rm{gs}}/k_B$ of a few mK for $w=1/2$ (i.e., quadratic dispersion), in the absence of any spin bias (i.e., $\Delta \mu=0$).  In practice, for a good thermal contact at the interface, this corresponds to a temperature difference maintained across the magnon correlation length, which we estimate by the magnetic exchange length ($\sim10$ nm in YIG); such thermal gradients have already been realized in experiment \cite{UchidaAPL}.

Considering that the classically unstable region ($\Delta\mu>\epsilon_{\rm gs}$) has already been realized in practice \cite{kajiwaraNAT10} in a Pt/YIG bilayer spin-biased by the inverse spin Hall effect, and the spin-caloritronic properties \cite{bauerCM11} are presently under intense experimental scrutiny in such composites \cite{sandwegPRL11,uchidaNATM10}, the experimental observation of current-induced BEC phase in Pt/YIG hybrids appears very feasible. YIG film thickness larger than the characteristic de Broglie wavelength of magnons ($\sim1$~nm at room temperature using standard YIG parameters \cite{bhagatPSS73}) would justify a three-dimensional treatment of BEC. A $d_L\lesssim1$~$\mu$m-thick YIG film with Gilbert damping $\alpha\lesssim10^{-4}$ like that employed in Ref.~\cite{kajiwaraNAT10} appears adequate to our ends, in order for the spin-pumping efficiency $\alpha'$ to be comparable to the intrinsic Gilbert damping $\alpha$.

We conclude that BEC phase can be established under a steady-state transport condition when the ferromagnet is  colder than the normal metal (thus facilitated by a spin Seebeck effect \cite{bauerCM11}) and the spin accumulation $\Delta\mu$ is slightly below the spin-transfer torque instability ($\Delta\mu\sim\epsilon_{\rm gs}$), in our model. Implicit in our discussion is the assumption that the magnon gas is dilute and can therefore be treated as noninteracting, aside from thermalization effects.  In reality, these interactions must be accounted for, in order to fully understand the ensuing dynamics of the magnon condensate. In such treatment, spectral properties would be self-consistently modified deep in the BEC phase, but the essential behavior of the system close to the transition point could still be addressed by the present theory. The emergent magnon superfluid properties \cite{bunkovJPCM10} due to their interactions are left for a future work.

The authors would like to thank Silas Hoffman for his fruitful insights. This work was supported in part by the NSF under Grant No. DMR-0840965, DARPA (Y.T.), the FOM, NWO, and ERC (R.D.).


\begin{thebibliography}{22}%
\makeatletter
\providecommand \@ifxundefined [1]{%
 \@ifx{#1\undefined}
}%
\providecommand \@ifnum [1]{%
 \ifnum #1\expandafter \@firstoftwo
 \else \expandafter \@secondoftwo
 \fi
}%
\providecommand \@ifx [1]{%
 \ifx #1\expandafter \@firstoftwo
 \else \expandafter \@secondoftwo
 \fi
}%
\providecommand \natexlab [1]{#1}%
\providecommand \enquote  [1]{``#1''}%
\providecommand \bibnamefont  [1]{#1}%
\providecommand \bibfnamefont [1]{#1}%
\providecommand \citenamefont [1]{#1}%
\providecommand \href@noop [0]{\@secondoftwo}%
\providecommand \href [0]{\begingroup \@sanitize@url \@href}%
\providecommand \@href[1]{\@@startlink{#1}\@@href}%
\providecommand \@@href[1]{\endgroup#1\@@endlink}%
\providecommand \@sanitize@url [0]{\catcode `\\12\catcode `\$12\catcode
  `\&12\catcode `\#12\catcode `\^12\catcode `\_12\catcode `\%12\relax}%
\providecommand \@@startlink[1]{}%
\providecommand \@@endlink[0]{}%
\providecommand \url  [0]{\begingroup\@sanitize@url \@url }%
\providecommand \@url [1]{\endgroup\@href {#1}{\urlprefix }}%
\providecommand \urlprefix  [0]{URL }%
\providecommand \Eprint [0]{\href }%
\@ifxundefined \urlstyle {%
  \providecommand \doi  [0]{\begingroup \@sanitize@url \@doi}%
  \providecommand \@doi [1]{\endgroup \@@startlink {\doibase
  #1}doi:\discretionary {}{}{}#1\@@endlink }%
}{%
  \providecommand \doi  [0]{doi:\discretionary{}{}{}\begingroup
  \urlstyle{rm}\Url }%
}%
\providecommand \doibase [0]{http://dx.doi.org/}%
\providecommand \Doi [0]{\begingroup \@sanitize@url \@Doi }%
\providecommand \@Doi  [1]{\endgroup\@@startlink{\doibase#1}\@@Doi}%
\providecommand \@@Doi [1]{#1\@@endlink}%
\providecommand \selectlanguage [0]{\@gobble}%
\providecommand \bibinfo  [0]{\@secondoftwo}%
\providecommand \bibfield  [0]{\@secondoftwo}%
\providecommand \translation [1]{[#1]}%
\providecommand \BibitemOpen [0]{}%
\providecommand \bibitemStop [0]{}%
\providecommand \bibitemNoStop [0]{.\EOS\space}%
\providecommand \EOS [0]{\spacefactor3000\relax}%
\providecommand \BibitemShut  [1]{\csname bibitem#1\endcsname}%
\bibitem [{\citenamefont {Anderson}\ \emph {et~al.}(1995)\citenamefont
  {Anderson}, \citenamefont {Ensher}, \citenamefont {Matthews}, \citenamefont
  {Wieman},\ and\ \citenamefont {Cornell}}]{andersonSCI95}%
  \BibitemOpen
  \bibfield  {author} {\bibinfo {author} {\bibfnamefont {M.~H.}\ \bibnamefont
  {Anderson}}, \bibinfo {author} {\bibfnamefont {J.~R.}\ \bibnamefont
  {Ensher}}, \bibinfo {author} {\bibfnamefont {M.~R.}\ \bibnamefont
  {Matthews}}, \bibinfo {author} {\bibfnamefont {C.~E.}\ \bibnamefont
  {Wieman}}, \ and\ \bibinfo {author} {\bibfnamefont {E.~A.}\ \bibnamefont
  {Cornell}},\ }\href@noop {} {\bibfield  {journal} {\bibinfo  {journal}
  {Science},\ }\textbf {\bibinfo {volume} {269}},\ \bibinfo {pages} {198}
  (\bibinfo {year} {1995})}\BibitemShut {NoStop}%
\bibitem [{\citenamefont {Mewes}\ \emph {et~al.}(1996)\citenamefont {Mewes},
  \citenamefont {Andrews}, \citenamefont {{van Druten}}, \citenamefont {Kurn},
  \citenamefont {Durfee},\ and\ \citenamefont {Ketterle}}]{mewesPRL96}%
  \BibitemOpen
  \bibfield  {author} {\bibinfo {author} {\bibfnamefont {M.-O.}\ \bibnamefont
  {Mewes}}, \bibinfo {author} {\bibfnamefont {M.~R.}\ \bibnamefont {Andrews}},
  \bibinfo {author} {\bibfnamefont {N.~J.}\ \bibnamefont {{van Druten}}},
  \bibinfo {author} {\bibfnamefont {D.~M.}\ \bibnamefont {Kurn}}, \bibinfo
  {author} {\bibfnamefont {D.~S.}\ \bibnamefont {Durfee}}, \ and\ \bibinfo
  {author} {\bibfnamefont {W.}~\bibnamefont {Ketterle}},\ }\Doi
  {10.1103/PhysRevLett.77.416} {\bibfield  {journal} {\bibinfo  {journal}
  {Phys. Rev. Lett.},\ }\textbf {\bibinfo {volume} {77}},\ \bibinfo {pages}
  {416} (\bibinfo {year} {1996})}\BibitemShut {NoStop}%
\bibitem [{\citenamefont {Zwierlein}\ \emph {et~al.}(2003)\citenamefont
  {Zwierlein}, \citenamefont {Stan}, \citenamefont {Schunck}, \citenamefont
  {Raupach}, \citenamefont {Gupta}, \citenamefont {Hadzibabic},\ and\
  \citenamefont {Ketterle}}]{zwierleinPRL03}%
  \BibitemOpen
  \bibfield  {author} {\bibinfo {author} {\bibfnamefont {M.~W.}\ \bibnamefont
  {Zwierlein}}, \bibinfo {author} {\bibfnamefont {C.~A.}\ \bibnamefont {Stan}},
  \bibinfo {author} {\bibfnamefont {C.~H.}\ \bibnamefont {Schunck}}, \bibinfo
  {author} {\bibfnamefont {S.~M.~F.}\ \bibnamefont {Raupach}}, \bibinfo
  {author} {\bibfnamefont {S.}~\bibnamefont {Gupta}}, \bibinfo {author}
  {\bibfnamefont {Z.}~\bibnamefont {Hadzibabic}}, \ and\ \bibinfo {author}
  {\bibfnamefont {W.}~\bibnamefont {Ketterle}},\ }\Doi
  {10.1103/PhysRevLett.91.250401} {\bibfield  {journal} {\bibinfo  {journal}
  {Phys. Rev. Lett.},\ }\textbf {\bibinfo {volume} {91}},\ \bibinfo {pages}
  {250401} (\bibinfo {year} {2003})}\BibitemShut {NoStop}%
\bibitem [{\citenamefont {Zwierlein}\ \emph {et~al.}(2006)\citenamefont
  {Zwierlein}, \citenamefont {Schirotzek}, \citenamefont {Schunck},\ and\
  \citenamefont {Ketterle}}]{zwierleinSCI06}%
  \BibitemOpen
  \bibfield  {author} {\bibinfo {author} {\bibfnamefont {M.~W.}\ \bibnamefont
  {Zwierlein}}, \bibinfo {author} {\bibfnamefont {A.}~\bibnamefont
  {Schirotzek}}, \bibinfo {author} {\bibfnamefont {C.~H.}\ \bibnamefont
  {Schunck}}, \ and\ \bibinfo {author} {\bibfnamefont {W.}~\bibnamefont
  {Ketterle}},\ }\href@noop {} {\bibfield  {journal} {\bibinfo  {journal}
  {Science},\ }\textbf {\bibinfo {volume} {311}},\ \bibinfo {pages} {492}
  (\bibinfo {year} {2006})}\BibitemShut {NoStop}%
\bibitem [{\citenamefont {Deng}\ \emph {et~al.}(2002)\citenamefont {Deng},
  \citenamefont {Weihs}, \citenamefont {Santori}, \citenamefont {Bloch},\ and\
  \citenamefont {Yamamoto}}]{dengSCI02}%
  \BibitemOpen
  \bibfield  {author} {\bibinfo {author} {\bibfnamefont {H.}~\bibnamefont
  {Deng}}, \bibinfo {author} {\bibfnamefont {G.}~\bibnamefont {Weihs}},
  \bibinfo {author} {\bibfnamefont {C.}~\bibnamefont {Santori}}, \bibinfo
  {author} {\bibfnamefont {J.}~\bibnamefont {Bloch}}, \ and\ \bibinfo {author}
  {\bibfnamefont {Y.}~\bibnamefont {Yamamoto}},\ }\href@noop {} {\bibfield
  {journal} {\bibinfo  {journal} {Science},\ }\textbf {\bibinfo {volume}
  {298}},\ \bibinfo {pages} {199} (\bibinfo {year} {2002})}\BibitemShut
  {NoStop}%
\bibitem [{\citenamefont {Kasprzak}\ \emph {et~al.}(2006)\citenamefont
  {Kasprzak}, \citenamefont {Richard}, \citenamefont {Kundermann},
  \citenamefont {Baas}, \citenamefont {Jeambrun}, \citenamefont {Keeling},
  \citenamefont {Marchetti}, \citenamefont {Szyma{\'n}ska}, \citenamefont
  {Andr{\'e}}, \citenamefont {Staehli}, \citenamefont {Savona}, \citenamefont
  {Littlewood}, \citenamefont {Deveaud},\ and\ \citenamefont
  {Dang}}]{kasprzakNAT06}%
  \BibitemOpen
  \bibfield  {author} {\bibinfo {author} {\bibfnamefont {J.}~\bibnamefont
  {Kasprzak}}, \bibinfo {author} {\bibfnamefont {M.}~\bibnamefont {Richard}},
  \bibinfo {author} {\bibfnamefont {S.}~\bibnamefont {Kundermann}}, \bibinfo
  {author} {\bibfnamefont {A.}~\bibnamefont {Baas}}, \bibinfo {author}
  {\bibfnamefont {P.}~\bibnamefont {Jeambrun}}, \bibinfo {author}
  {\bibfnamefont {J.~M.~J.}\ \bibnamefont {Keeling}}, \bibinfo {author}
  {\bibfnamefont {F.~M.}\ \bibnamefont {Marchetti}}, \bibinfo {author}
  {\bibfnamefont {M.~H.}\ \bibnamefont {Szyma{\'n}ska}}, \bibinfo {author}
  {\bibfnamefont {R.}~\bibnamefont {Andr{\'e}}}, \bibinfo {author}
  {\bibfnamefont {J.~L.}\ \bibnamefont {Staehli}}, \bibinfo {author}
  {\bibfnamefont {V.}~\bibnamefont {Savona}}, \bibinfo {author} {\bibfnamefont
  {P.~B.}\ \bibnamefont {Littlewood}}, \bibinfo {author} {\bibfnamefont
  {B.}~\bibnamefont {Deveaud}}, \ and\ \bibinfo {author} {\bibfnamefont
  {L.~S.}\ \bibnamefont {Dang}},\ }\href@noop {} {\bibfield  {journal}
  {\bibinfo  {journal} {Nature},\ }\textbf {\bibinfo {volume} {443}},\ \bibinfo
  {pages} {409} (\bibinfo {year} {2006})}\BibitemShut {NoStop}%
\bibitem [{\citenamefont {Balili}\ \emph {et~al.}(2007)\citenamefont {Balili},
  \citenamefont {Hartwell}, \citenamefont {Snoke}, \citenamefont {Pfeiffer},\
  and\ \citenamefont {West}}]{baliliSCI07}%
  \BibitemOpen
  \bibfield  {author} {\bibinfo {author} {\bibfnamefont {R.}~\bibnamefont
  {Balili}}, \bibinfo {author} {\bibfnamefont {V.}~\bibnamefont {Hartwell}},
  \bibinfo {author} {\bibfnamefont {D.}~\bibnamefont {Snoke}}, \bibinfo
  {author} {\bibfnamefont {L.}~\bibnamefont {Pfeiffer}}, \ and\ \bibinfo
  {author} {\bibfnamefont {K.}~\bibnamefont {West}},\ }\href@noop {} {\bibfield
   {journal} {\bibinfo  {journal} {Science},\ }\textbf {\bibinfo {volume}
  {316}},\ \bibinfo {pages} {1007} (\bibinfo {year} {2007})}\BibitemShut
  {NoStop}%
\bibitem [{\citenamefont {Klaers}\ \emph {et~al.}(2010)\citenamefont {Klaers},
  \citenamefont {Schmitt}, \citenamefont {Vewinger},\ and\ \citenamefont
  {Weitz}}]{klaersNAT10}%
  \BibitemOpen
  \bibfield  {author} {\bibinfo {author} {\bibfnamefont {J.}~\bibnamefont
  {Klaers}}, \bibinfo {author} {\bibfnamefont {J.}~\bibnamefont {Schmitt}},
  \bibinfo {author} {\bibfnamefont {F.}~\bibnamefont {Vewinger}}, \ and\
  \bibinfo {author} {\bibfnamefont {M.}~\bibnamefont {Weitz}},\ }\href@noop {}
  {\bibfield  {journal} {\bibinfo  {journal} {Nature},\ }\textbf {\bibinfo
  {volume} {468}},\ \bibinfo {pages} {545} (\bibinfo {year}
  {2010})}\BibitemShut {NoStop}%
\bibitem [{\citenamefont {Demokritov}\ \emph {et~al.}(2006)\citenamefont
  {Demokritov}, \citenamefont {Demidov}, \citenamefont {Dzyapko}, \citenamefont
  {Melkov}, \citenamefont {Serga}, \citenamefont {Hillebrands},\ and\
  \citenamefont {Slavin}}]{demokritovNAT06}%
  \BibitemOpen
  \bibfield  {author} {\bibinfo {author} {\bibfnamefont {S.~O.}\ \bibnamefont
  {Demokritov}}, \bibinfo {author} {\bibfnamefont {V.~E.}\ \bibnamefont
  {Demidov}}, \bibinfo {author} {\bibfnamefont {O.}~\bibnamefont {Dzyapko}},
  \bibinfo {author} {\bibfnamefont {G.~A.}\ \bibnamefont {Melkov}}, \bibinfo
  {author} {\bibfnamefont {A.~A.}\ \bibnamefont {Serga}}, \bibinfo {author}
  {\bibfnamefont {B.}~\bibnamefont {Hillebrands}}, \ and\ \bibinfo {author}
  {\bibfnamefont {A.~N.}\ \bibnamefont {Slavin}},\ }\href@noop {} {\bibfield
  {journal} {\bibinfo  {journal} {Nature},\ }\textbf {\bibinfo {volume}
  {443}},\ \bibinfo {pages} {430} (\bibinfo {year} {2006})}\BibitemShut
  {NoStop}%
\bibitem [{\citenamefont {Demidov}\ \emph {et~al.}(2008)\citenamefont
  {Demidov}, \citenamefont {Dzyapko}, \citenamefont {Demokritov}, \citenamefont
  {Melkov},\ and\ \citenamefont {Slavin}}]{demidovPRL08}%
  \BibitemOpen
  \bibfield  {author} {\bibinfo {author} {\bibfnamefont {V.~E.}\ \bibnamefont
  {Demidov}}, \bibinfo {author} {\bibfnamefont {O.}~\bibnamefont {Dzyapko}},
  \bibinfo {author} {\bibfnamefont {S.~O.}\ \bibnamefont {Demokritov}},
  \bibinfo {author} {\bibfnamefont {G.~A.}\ \bibnamefont {Melkov}}, \ and\
  \bibinfo {author} {\bibfnamefont {A.~N.}\ \bibnamefont {Slavin}},\ }\Doi
  {10.1103/PhysRevLett.100.047205} {\bibfield  {journal} {\bibinfo  {journal}
  {Phys. Rev. Lett.},\ }\textbf {\bibinfo {volume} {100}},\ \bibinfo {pages}
  {047205} (\bibinfo {year} {2008})}\BibitemShut {NoStop}%
\bibitem [{\citenamefont {Snoke}(2006)}]{snokeNAT06}%
  \BibitemOpen
  \bibfield  {author} {\bibinfo {author} {\bibfnamefont {D.}~\bibnamefont
  {Snoke}},\ }\href@noop {} {\bibfield  {journal} {\bibinfo  {journal}
  {Nature},\ }\textbf {\bibinfo {volume} {443}},\ \bibinfo {pages} {403}
  (\bibinfo {year} {2006})}\BibitemShut {NoStop}%
\bibitem [{\citenamefont {Tserkovnyak}\ \emph {et~al.}(2002)\citenamefont
  {Tserkovnyak}, \citenamefont {Brataas},\ and\ \citenamefont
  {Bauer}}]{tserkovPRL02sp}%
  \BibitemOpen
  \bibfield  {author} {\bibinfo {author} {\bibfnamefont {Y.}~\bibnamefont
  {Tserkovnyak}}, \bibinfo {author} {\bibfnamefont {A.}~\bibnamefont
  {Brataas}}, \ and\ \bibinfo {author} {\bibfnamefont {G.~E.~W.}\ \bibnamefont
  {Bauer}},\ }\href@noop {} {\bibfield  {journal} {\bibinfo  {journal} {Phys.
  Rev. Lett.},\ }\textbf {\bibinfo {volume} {88}},\ \bibinfo {eid} {117601}
  (\bibinfo {year} {2002})}\BibitemShut {NoStop}%
\bibitem [{\citenamefont {Tserkovnyak}\ \emph {et~al.}(2005)\citenamefont
  {Tserkovnyak}, \citenamefont {Brataas}, \citenamefont {Bauer},\ and\
  \citenamefont {Halperin}}]{tserkovRMP05}%
  \BibitemOpen
  \bibfield  {author} {\bibinfo {author} {\bibfnamefont {Y.}~\bibnamefont
  {Tserkovnyak}}, \bibinfo {author} {\bibfnamefont {A.}~\bibnamefont
  {Brataas}}, \bibinfo {author} {\bibfnamefont {G.~E.~W.}\ \bibnamefont
  {Bauer}}, \ and\ \bibinfo {author} {\bibfnamefont {B.~I.}\ \bibnamefont
  {Halperin}},\ }\href@noop {} {\bibfield  {journal} {\bibinfo  {journal} {Rev.
  Mod. Phys.},\ }\textbf {\bibinfo {volume} {77}},\ \bibinfo {eid} {1375}
  (\bibinfo {year} {2005})}\BibitemShut {NoStop}%
\bibitem [{\citenamefont {Slonczewski}(1996)}]{slonczewskiJMMM96}%
  \BibitemOpen
  \bibfield  {author} {\bibinfo {author} {\bibfnamefont {J.~C.}\ \bibnamefont
  {Slonczewski}},\ }\href@noop {} {\bibfield  {journal} {\bibinfo  {journal}
  {J. Magn. Magn. Mater.},\ }\textbf {\bibinfo {volume} {159}},\ \bibinfo
  {pages} {L1} (\bibinfo {year} {1996})}\BibitemShut {NoStop}%
\bibitem [{\citenamefont {Berger}(1996)}]{bergerPRB96}%
  \BibitemOpen
  \bibfield  {author} {\bibinfo {author} {\bibfnamefont {L.}~\bibnamefont
  {Berger}},\ }\href@noop {} {\bibfield  {journal} {\bibinfo  {journal} {Phys.
  Rev. B},\ }\textbf {\bibinfo {volume} {54}},\ \bibinfo {pages} {9353}
  (\bibinfo {year} {1996})}\BibitemShut {NoStop}%
\bibitem [{\citenamefont {Bauer}\ and\ \citenamefont
  {Tserkovnyak}(2011)}]{bauerPHYS11}%
  \BibitemOpen
  \bibfield  {author} {\bibinfo {author} {\bibfnamefont {G.~E.~W.}\
  \bibnamefont {Bauer}}\ and\ \bibinfo {author} {\bibfnamefont
  {Y.}~\bibnamefont {Tserkovnyak}},\ }\href@noop {} {\bibfield  {journal}
  {\bibinfo  {journal} {Physics},\ }\textbf {\bibinfo {volume} {4}},\ \bibinfo
  {pages} {40} (\bibinfo {year} {2011})}\BibitemShut {NoStop}%
\bibitem [{\citenamefont {Sandweg}\ \emph {et~al.}(2011)\citenamefont
  {Sandweg}, \citenamefont {Kajiwara}, \citenamefont {Chumak}, \citenamefont
  {Serga}, \citenamefont {Vasyuchka}, \citenamefont {Jungfleisch},
  \citenamefont {Saitoh},\ and\ \citenamefont {Hillebrands}}]{sandwegPRL11}%
  \BibitemOpen
  \bibfield  {author} {\bibinfo {author} {\bibfnamefont {C.~W.}\ \bibnamefont
  {Sandweg}}, \bibinfo {author} {\bibfnamefont {Y.}~\bibnamefont {Kajiwara}},
  \bibinfo {author} {\bibfnamefont {A.~V.}\ \bibnamefont {Chumak}}, \bibinfo
  {author} {\bibfnamefont {A.~A.}\ \bibnamefont {Serga}}, \bibinfo {author}
  {\bibfnamefont {V.~I.}\ \bibnamefont {Vasyuchka}}, \bibinfo {author}
  {\bibfnamefont {M.~B.}\ \bibnamefont {Jungfleisch}}, \bibinfo {author}
  {\bibfnamefont {E.}~\bibnamefont {Saitoh}}, \ and\ \bibinfo {author}
  {\bibfnamefont {B.}~\bibnamefont {Hillebrands}},\ }\Doi
  {10.1103/PhysRevLett.106.216601} {\bibfield  {journal} {\bibinfo  {journal}
  {Phys. Rev. Lett.},\ }\textbf {\bibinfo {volume} {106}},\ \bibinfo {pages}
  {216601} (\bibinfo {year} {2011})}\BibitemShut {NoStop}%
\bibitem [{\citenamefont {Kajiwara}\ \emph {et~al.}(2010)\citenamefont
  {Kajiwara}, \citenamefont {Harii}, \citenamefont {Takahashi}, \citenamefont
  {Ohe}, \citenamefont {Uchida}, \citenamefont {Mizuguchi}, \citenamefont
  {Umezawa}, \citenamefont {Kawai}, \citenamefont {Ando}, \citenamefont
  {Takanashi}, \citenamefont {Maekawa},\ and\ \citenamefont
  {Saitoh}}]{kajiwaraNAT10}%
  \BibitemOpen
  \bibfield  {author} {\bibinfo {author} {\bibfnamefont {Y.}~\bibnamefont
  {Kajiwara}}, \bibinfo {author} {\bibfnamefont {K.}~\bibnamefont {Harii}},
  \bibinfo {author} {\bibfnamefont {S.}~\bibnamefont {Takahashi}}, \bibinfo
  {author} {\bibfnamefont {J.}~\bibnamefont {Ohe}}, \bibinfo {author}
  {\bibfnamefont {K.}~\bibnamefont {Uchida}}, \bibinfo {author} {\bibfnamefont
  {M.}~\bibnamefont {Mizuguchi}}, \bibinfo {author} {\bibfnamefont
  {H.}~\bibnamefont {Umezawa}}, \bibinfo {author} {\bibfnamefont
  {H.}~\bibnamefont {Kawai}}, \bibinfo {author} {\bibfnamefont
  {K.}~\bibnamefont {Ando}}, \bibinfo {author} {\bibfnamefont {K.}~\bibnamefont
  {Takanashi}}, \bibinfo {author} {\bibfnamefont {S.}~\bibnamefont {Maekawa}},
  \ and\ \bibinfo {author} {\bibfnamefont {E.}~\bibnamefont {Saitoh}},\
  }\href@noop {} {\bibfield  {journal} {\bibinfo  {journal} {Nature},\ }\textbf
  {\bibinfo {volume} {464}},\ \bibinfo {pages} {262} (\bibinfo {year}
  {2010})}\BibitemShut {NoStop}%
\bibitem [{\citenamefont {Bauer}()}]{bauerCM11}%
  \BibitemOpen
  \bibfield  {author} {\bibinfo {author} {\bibfnamefont {G.~E.~W.}\
  \bibnamefont {Bauer}},\ }\href@noop {} {\enquote {\bibinfo {title} {Spin
  caloritronics},}\ }\bibinfo {note} {arXiv:1107.4395}\BibitemShut {NoStop}%
\bibitem [{\citenamefont {Heinrich}\ \emph {et~al.}(2011)\citenamefont {Heinrich},
  \citenamefont {Burrowes}, \citenamefont {Montoya}, \citenamefont {Kardasz}, \citenamefont {Girt}, \citenamefont {Song}, \citenamefont {Sun},\ and\ \citenamefont {Wu}}]{heinrichPRL11}%
  \BibitemOpen
  \bibfield  {author} {\bibinfo {author} {\bibfnamefont {B.}~\bibnamefont
  {Heinrich}}, \bibinfo {author} {\bibfnamefont {C.}~\bibnamefont {Burrowes}},
  \bibinfo {author} {\bibfnamefont {E.}~\bibnamefont {Montoya}},
  \bibinfo {author} {\bibfnamefont {B.}~\bibnamefont {Kardasz}},
  \bibinfo {author} {\bibfnamefont {E.}~\bibnamefont {Girt}},
  \bibinfo {author} {\bibfnamefont {Y.-Y.}~\bibnamefont {Song}},
  \bibinfo {author} {\bibfnamefont {Y.}~\bibnamefont {Sun}},\ and\
  \bibinfo {author} {\bibfnamefont {M.}~\bibnamefont {Wu}},\ }\href@noop
  {} {\bibfield  {journal} {\bibinfo  {journal} {Phys. Rev. Lett.},\
  }\textbf {\bibinfo {volume} {107}},\ \bibinfo {pages} {066604} (\bibinfo {year}
  {2011})}\BibitemShut {NoStop}%
\bibitem{jia} X.~Jia, K.~Liu, K.~Xia, and G.~E.~W. Bauer, Europhys. Lett. {\bf 96}, 17005 (2011).
\bibitem{burrows} C.~Burrowes, B.~Heinrich, B.~Kardasz, E.~A. Montoya, E.~Girt, Y. Sun, Y.-Y. Song, and M. Wu, unpublished.

\bibitem{UchidaAPL} K.~Uchida, H.~Adachi, T.~Ota, H.~Nakayama, S.~W. Maekawa, and E.~Saitoh, App. Phys. Lett.  {\bf 97}, 172505 (2010).

\bibitem [{\citenamefont {Uchida}\ \emph {et~al.}(2010)\citenamefont {Uchida},
  \citenamefont {Xiao}, \citenamefont {Adachi}, \citenamefont {Ohe},
  \citenamefont {Takahashi}, \citenamefont {Ieda}, \citenamefont {Ota},
  \citenamefont {Kajiwara}, \citenamefont {Umezawa}, \citenamefont {Kawai},
  \citenamefont {Bauer}, \citenamefont {Maekawa},\ and\ \citenamefont
  {Saitoh}}]{uchidaNATM10}%
  \BibitemOpen
  \bibfield  {author} {\bibinfo {author} {\bibfnamefont {K.}~\bibnamefont
  {Uchida}}, \bibinfo {author} {\bibfnamefont {J.}~\bibnamefont {Xiao}},
  \bibinfo {author} {\bibfnamefont {H.}~\bibnamefont {Adachi}}, \bibinfo
  {author} {\bibfnamefont {J.}~\bibnamefont {Ohe}}, \bibinfo {author}
  {\bibfnamefont {S.}~\bibnamefont {Takahashi}}, \bibinfo {author}
  {\bibfnamefont {J.}~\bibnamefont {Ieda}}, \bibinfo {author} {\bibfnamefont
  {T.}~\bibnamefont {Ota}}, \bibinfo {author} {\bibfnamefont {Y.}~\bibnamefont
  {Kajiwara}}, \bibinfo {author} {\bibfnamefont {H.}~\bibnamefont {Umezawa}},
  \bibinfo {author} {\bibfnamefont {H.}~\bibnamefont {Kawai}}, \bibinfo
  {author} {\bibfnamefont {G.~E.~W.}\ \bibnamefont {Bauer}}, \bibinfo {author}
  {\bibfnamefont {S.}~\bibnamefont {Maekawa}}, \ and\ \bibinfo {author}
  {\bibfnamefont {E.}~\bibnamefont {Saitoh}},\ }\href@noop {} {\bibfield
  {journal} {\bibinfo  {journal} {Nature Mater.},\ }\textbf {\bibinfo {volume}
  {9}},\ \bibinfo {pages} {894} (\bibinfo {year} {2010})}\BibitemShut {NoStop}%
\bibitem [{\citenamefont {Bhagat}\ \emph {et~al.}(1973)\citenamefont {Bhagat},
  \citenamefont {Lessoff}, \citenamefont {Vittoria},\ and\ \citenamefont
  {Guenzer}}]{bhagatPSS73}%
  \BibitemOpen
  \bibfield  {author} {\bibinfo {author} {\bibfnamefont {S.}~\bibnamefont
  {Bhagat}}, \bibinfo {author} {\bibfnamefont {H.}~\bibnamefont {Lessoff}},
  \bibinfo {author} {\bibfnamefont {C.}~\bibnamefont {Vittoria}}, \ and\
  \bibinfo {author} {\bibfnamefont {C.}~\bibnamefont {Guenzer}},\ }\href@noop
  {} {\bibfield  {journal} {\bibinfo  {journal} {Phys. Status Solidi},\
  }\textbf {\bibinfo {volume} {20}},\ \bibinfo {pages} {731} (\bibinfo {year}
  {1973})}\BibitemShut {NoStop}%
\bibitem [{\citenamefont {Bunkov}\ and\ \citenamefont
  {Volovik}(2010)}]{bunkovJPCM10}%
  \BibitemOpen
  \bibfield  {author} {\bibinfo {author} {\bibfnamefont {Y.~M.}\ \bibnamefont
  {Bunkov}}\ and\ \bibinfo {author} {\bibfnamefont {G.~E.}\ \bibnamefont
  {Volovik}},\ }\href@noop {} {\bibfield  {journal} {\bibinfo  {journal} {J.
  Phys.: Condens. Matter},\ }\textbf {\bibinfo {volume} {22}},\ \bibinfo
  {pages} {164210} (\bibinfo {year} {2010})}\BibitemShut {NoStop}%
\end{thebibliography}
\end{document}


\section{Thermodynamics of spin transport}

Eq.~(6) of the main text suggests that spin flow to and from excited magnon states vanishes when there is no thermal or spin gradient, i.e., when $\beta_{L}=\beta_{R}$ and $\Delta\mu=\mu_{L}$.
However, when either of these conditions is not met, $j_{\rm ex}\neq0$
and spin (as well as energy) is transported across the insulator/conductor interface. In a steady state (i.e., zero spin current), in normal phase with thermal bias, $\beta_{R}-\beta_{L}\neq0$, a spin chemical potential difference $\delta\mu=\Delta\mu-\mu_{L}$ develops to oppose it:
\[
\delta\mu \approx\left(\beta_{R}-\beta_{L}\right)\frac{\int_{\epsilon_{\rm gs}}^{\infty} d\epsilon\left( \epsilon-\bar{\mu} \right) \left(\epsilon-\Delta\mu\right)n'_{\mathrm{B}}\left(\bar{\beta}\left(\epsilon-\bar{\mu}\right) \right)}{\bar{\beta} \int_{\epsilon_{\rm gs}}^{\infty} d\epsilon  \left(\epsilon-\Delta\mu\right)n'_{\mathrm{B}}\left(\bar{\beta}\left(\epsilon -\bar{\mu}\right) \right)}\,,\]
where $\bar{\beta}\equiv \left(\beta_{L}+\beta_{R} \right)/2$, $\bar{\mu}\equiv\left(\mu_L+\Delta \mu \right)/2$, $n'_B=\partial_\epsilon n_B (\epsilon)$,  and the thermodynamic biases are assumed to be small (i.e. $ \beta_R-\beta_L \ll \bar{\beta},  \delta \mu\ll\bar{\mu}$). 

\begin{figure}[h]
\includegraphics[width=0.55\linewidth,clip=]{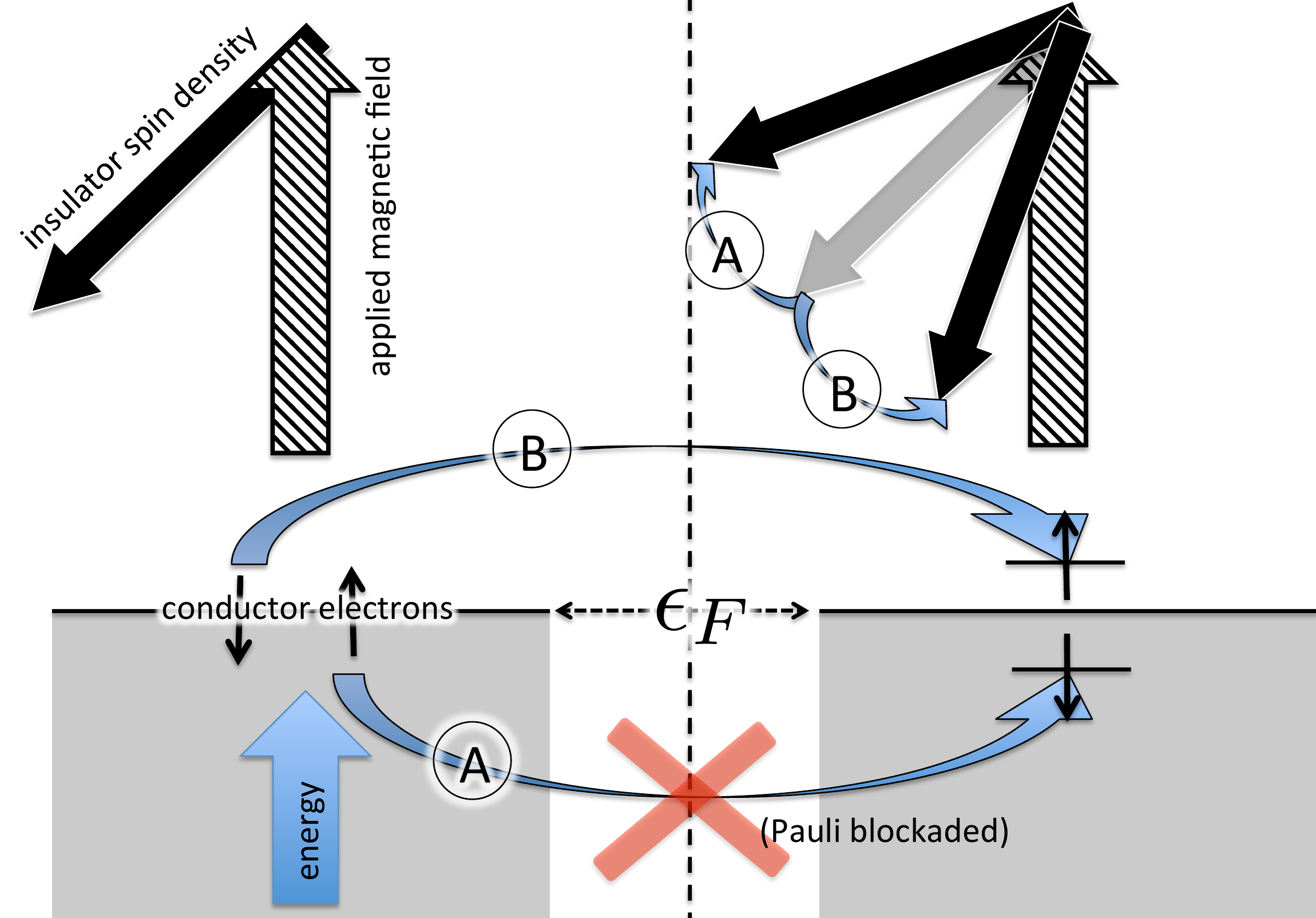}
\caption{A down electron may relax the ferromagnetic insulator, carrying away the excess energy away in a scattering state above the Fermi surface $\epsilon_{F}$ (process B). An incident up electron on the Fermi surface, however, cannot transfer up spin to the insulator magnetization (process A), since such an energy-preserving process would raise the energy of the magnet, lowering that of the electron and therefore landing it below the Fermi surface, which is Pauli blockaded. Process B therefore dominates, and the insulator magnetization relaxes towards the easy axis.}
\label{Excluded}
\end{figure}

On the other hand, the condensed spin current $j_{\rm gs}$
is independent of both $T_L$ and $T_R$, and, provided $\epsilon_{\rm gs}>\Delta\mu$,
always carries spin away from the conductor, irrespective of the temperature
gradient between the two systems. The explanation for this behavior
can be understood as follows. Consider a single tunneling event involving
the creation (destruction) of a ground-state magnon ($\Delta N_{{\rm gs}}=\pm1$)
and the corresponding creation of a down-(up-)spin electron-hole excitation
in the conductor ($\Delta N_{R}=-\Delta N_{\rm gs}$), which we call process A (B) in Fig.~\ref{Excluded}. The entropy change in the insulator
associated with either process vanishes when the magnons form a BEC,
so that the entropy change of the whole system is just $dS_{R}$, which
can be found by enforcing energy conservation:\[
\Delta S_{\mathrm{tot}}=\Delta S_{R}=\frac{1}{T_{R}}\left(\epsilon_{\rm gs}-\Delta\mu\right)\Delta N_{R}\,.\]
Thus,  process B (A) is favored ($\Delta N_{R}\gtrless0$)  for tunneling events involving ground-state
magnons when $\epsilon_{\rm gs}\gtrless \Delta\mu$, in agreement with Eq.~(4) of the main text. Put differently, if $\Delta\mu=0$ the phase space of the conductor is unaffected with either the introduction of an up-spin excitation or the introduction of a down-spin excitation. However, process A requires the conductor to surrender an energy quantum $\epsilon_{\mathrm{gs}}$ to the insulator, whereas process B means a net gain in energy for the conductor; the overall entropy gain in the conductor (and therefore the entire system) is thus greater for process B than A.  The zero-temperature version of this explanation is presented graphically in Fig.~\ref{Excluded}.

\section{Detection of phase transition}

The BEC-normal phase transition presents some of the most interesting
physics of the system, yet as can be seen from Fig.~3 of the main text, it
is difficult to discern from the total magnetization of the insulator alone: Whereas for fixed $T_L$ the density of excited magnons $n_{\rm ex}$ plateaus as $z_L\rightarrow1$, the rate of change of the total number of magnons $\dot{n}_L=\dot{n}_{\rm ex}+\dot{n}_{\rm gs}$ remains always continuous function of time. The transition can, however, be observed by Brillioun
light scattering, wherein the scattered light intensity scales quadratically with the lateral junction size if the ground-state condensate is indeed coherent (see Ref.~[4] of the main text).

Alternatively, electron spin resonance (or, for that matter, any spectroscopic probe of a coherent microwave radiation) can provide clear evidence of the presence of quasiequilibrated Bose-Einstein condensation of magnons.  Consider a test-particle electron at a fixed distance $\mathbf{r}$ from the magnetic insulator.  Provided that the electron experiences the insulator as a single quantum magnetic moment $\hat{\mathbf{m}}$, one may neglect details involving spatial fluctuations of the magnetization and allow the two systems to interact via dipole-dipole coupling; the Hamiltonian describing the interaction is therefore of the form:
\[\hat{H}_{\rm d-d}=\sum_{i,j=x,y,z}\hat{m}_{i} T_{ij}\hat{\sigma}_{j}\,,\]
where $T_{ij}$ is a tensor that depends on $\mathbf{r}$ and $\hat{\sigma}$  is the electron spin operator.  Supposing the electron, subjected to a strong applied magnetic field in the $z$ direction, begins in the state $\left|\uparrow\right\rangle$, the probability that quantum fluctuations in the magnetization $\hat{\mathbf{m}}$ spin flip the electron is, to lowest order in $T_{ij}$,
\[P_{\uparrow\rightarrow\downarrow}(t)=\int_{0}^{t}dt'\int_{0}^{t}dt''\sum_{iji'j'}T_{ij}T_{i'j'}\left\langle \hat{m}_{i}\left(t'\right)\hat{m}_{i'}\left(t''\right)\right\rangle\left\langle \uparrow\right|\hat{\sigma}_{j}\left|\downarrow\right\rangle \left\langle \downarrow\right|\hat{\sigma}_{j'}\left|\uparrow\right\rangle e^{i\omega_{z}\left(t'-t''\right)}\,,\]
where $\omega_z$ is the electronic Larmor frequency in the applied magnetic field. Choosing our coordinate system to coincide with the eigenbasis of $T_{ij}$ and for simplicity asserting cylindrical symmetry around the $z$ axis (so that $T_{xx}=T_{yy}=T_{\perp}$), the transition probability becomes
\[P_{\uparrow\rightarrow\downarrow}\left(t\right)=\int_{0}^{t}dt'\int_{0}^{t}dt''T_{\perp}^2 \left\langle \hat{m}^{-}\left(t'\right)\hat{m}^{+}\left(t''\right)\right\rangle e^{i\omega_{z}\left(t'-t''\right)}=tT_\perp^2S_{-+}(\omega_z)\,,\]
where $S_{-+}(\omega)=\int dte^{i\omega t}\langle\hat{m}^-(t)\hat{m}^+(0)\rangle\propto N_{\rm gs}$ is the spectral density of magnetic oscillations in a steady state. The transition rate is thus proportional to $N_{\rm gs}$, which scales linearly with the lateral dimensions of our junction in BEC phase and is size independent in normal phase. This simple treatment is pertinent to the case when the magnons condense at $\mathbf{q}=0$. Otherwise (as is the case in YIG, for example) one needs to come up with means to couple coherently to magnetic fluctuations at a finite $\mathbf{q}$ (perhaps using some form of grating).